# Renormalization of quasiparticle band gap in doped two-dimensional materials from many-body calculations

Shiyuan Gao and Li Yang

Department of Physics and Institute of Materials Science and Engineering, Washington University in St. Louis, St. Louis, Missouri 63130, USA

## Abstract

Doped free carriers can substantially renormalize electronic self-energy and quasiparticle band gaps of two-dimensional (2D) materials. However, it is still challenging to quantitatively calculate this many-electron effect, particularly at the low doping density that is most relevant to realistic experiments and devices. Here we develop a first-principles-based effective-mass model within the GW approximation and show a dramatic band gap renormalization of a few hundred meV for typical 2D semiconductors. Moreover, we reveal the roles of different many-electron interactions: The Coulomb-hole contribution is dominant for low doping densities while the screened-exchange contribution is dominant for high doping densities. Three prototypical 2D materials are studied by this method, $h$-BN, $MoS_2$, and black phosphorus, covering insulators to semiconductors. Especially, anisotropic black phosphorus exhibits a surprisingly large band gap renormalization because of its smaller density-of-state that enhances the screened-exchange interactions. Our work demonstrates an efficient way to accurately calculate band gap renormalization and provides quantitative understanding of doping-dependent many-electron physics of general 2D semiconductors.

# I.      Introduction

The field of 2D materials has expanded greatly in the past few years, featuring a broad range of applications for electronic, photonic and piezoelectric devices [1-4], as well as exciting new physics to be realized, such as 2D ferroelectricity, ferromagnetism and exciton condensate [5-8]. Almost all the applications are premised on a good understanding the electronic properties of the material, especially the quasiparticle band gap. The *ab initio* GW method has been the most successful first-principles approach of calculating the quasiparticle band structure of bulk crystals as well as molecules and low-dimensional structures [9-12]. In particular, well-converged GW results in 2D crystals has been achieved recently as the accurate treatments to 2D screened Coulomb interaction were established [13-16]. However, much less is known about how doping, a common theme in the 2D semiconductors and its heterostructures [17-20], can affect the electronic structure.

Doped free carriers have several effects that are particularly enhanced on the electronic structure of low-dimensional materials. First, the large density-of-states (DOS) from the van Hove singularity magnifies the contribution from electron occupation. Second, the screening from doped free carriers has a stronger effect on lower-dimension structures because of the weaker intrinsic dielectric screening. Third, free carriers in low-dimensional systems form a low-energy acoustic plasmon which can dynamically couple with quasiparticles. These effects result in an enhanced many-body renormalization of quasiparticles energy, as shown from previous theoretical GW calculations in both semiconducting carbon nanotubes [21,22] and 2D transition metal dichalcogenides (TMDs) [23], and from experimental measurements [24-27]. More recently, beyond the nonlinear quasiparticle band gap renormalization of several hundred meV, the optical gap of monolayer TMDs was predicted to stay nearly constant due to a cancellation with the renormalization of exciton binding energy [28]. However, a complete picture of the quasiparticle renormalization within a wide range of doping density is not clear because of the limitation of k-point-grid-based first-principle method in resolving the low doping density, which is, however, the most essential for experiments and devices. Moreover, previous works and methods cannot be directly applied to studying several newly emerged 2D materials such as black phosphorus (BP) whose electronic structure is significantly anisotropic.

In this work, we have developed an effective mass model and applied asymptotic analysis to resolve band gap renormalization, using the GW approximation and the framework of previous work [23]. The effective mass model supplements the *ab initio* calculation by bridging the gap around low doping density. It reveals that the change of the dielectric screening, which appears in term of the Coulomb-hole self-energy, is the dominating contributing factors to the band gap renormalization at low doping density. The change in electron occupation, which appears in term of the screened-exchange self-energy, is more important at high doping density. Additionally, we study band gap renormalization of doped monolayer BP, where we generalize our method to systems with strong anisotropy, and show that the smaller DOS of BP near the band edge enhances the band gap renormalization at high doping density.

The rest of the paper is organized as follows: In Section II we layer down the theoretical framework of our approach, show the computational details, and discuss the materials' intrinsic properties. In Section III we construct our effective mass model of the GW self-energy and band gap renormalization of doped *h*-BN and MoS$_2$. In Section IV, we discuss band gap renormalization of monolayer BP, where our model is to be generalized to anisotropic systems. Finally, the main results will be summarized in Section V.

## II. Computational details and Intrinsic property

In this work, we choose three prototype monolayer 2D structures, including hexagonal BN (*h*-BN), 2H-phase MoS$_2$, and BP. They cover 2D materials from semiconductors to insulators and from isotropic ones to anisotropic ones. To study the effect of doping, we calculate the quasiparticle band structure of these materials from the first-principles density functional theory (DFT)+GW method. The DFT calculation serves as a mean-field starting point for the GW calculation. It is performed using the plane-wave pseudopotential method implemented in Quantum Espresso [29]. The generalized gradient approximation (GGA)-PBE exchange-correlation functional [30] is used along with a plane-wave cutoff of 90 Ry, 75 Ry and 35 Ry for *h*-BN, MoS$_2$, and BP, respectively. Doping is introduced by changing the total electron number with a compensating jellium background. This resembles the gate-tunable electrostatic doping commonly seen in 2D materials. Our calculation shows that doping has very little effects on the DFT eigenvalues and wavefunctions. This is not surprising because DFT is known for its deficiency at capturing many-electron effects that are, however, crucial for our studied band gap renormalization.

Beyond DFT, we employ the GW approximation to study quasiparticle energies. The self-energy in a doped material is expanded into three terms:

$$\Sigma = iGW = i(G_{int}W_{int} + \delta GW_{int} + G_{int}\delta W + \delta G\delta W) \equiv \Sigma_{int} + \Sigma_1 + \Sigma_2 + \Sigma_3 \qquad (1)$$

The first, "intrinsic" term ($\Sigma_{int}$) indicates the self-energy contribution coming from the intrinsic (undoped) system. The second term ($\Sigma_1$) is the self-energy correction due to the change of electron (hole) occupation alone under the intrinsic screening. The third term ($\Sigma_2$) is due to the change in screening, and the last term ($\Sigma_3$) is related to both factors. The calculation details of these doping-related terms will be discussed in the next section. As we will see, the dielectric screening $W = \epsilon^{-1}v$ and its change upon doping $\delta W$ play a central role in this band gap renormalization.

The intrinsic term ($\Sigma_{int}$) of the self-energy is calculated with the usual GW routine implemented in the BerkeleyGW package [11]. Truncated Coulomb interaction [31] is used along with sufficient vacuum to eliminate interactions between layers. The static dielectric function is calculated within the random phase approximation (RPA) with 8 Ry energy cutoff, 120 and 140 conduction bands, and $24 \times 24 \times 1$ and $28 \times 20 \times 1$ k-point grid respectively for *h*-BN and BP, which grants a converged band gap within 0.1 eV. For MoS$_2$, 10 Ry cutoff, 256 conduction bands and $24 \times 24 \times 1$ k-point grid is used. Although it has been shown that the true convergence of the band gap in MoS$_2$ would require a much larger number of bands and dielectric cutoff [15], as far as our main concern of band gap renormalization goes, this set of parameters is enough. This is because the doping effect is mainly concentrated on small **q** and head (**G**=**G'**=0) part of the dielectric function $\epsilon_{GG'}^{-1}(\boldsymbol{q},\omega)$ [23]. The dynamical part of the dielectric function is then constructed from the generalized plasmon-pole (GPP) model.

Figure 1 shows the calculated static dielectric function $\epsilon_{00}^{-1}(\boldsymbol{q},\omega=0)$ of intrinsic *h*-BN, MoS$_2$ and BP. The dielectric function approaches 1 in the as $q \to 0$, following the formula $\epsilon_{00}^{-1}(\boldsymbol{q}) \approx 1/(1+2\pi\alpha_{2D}q)$, where the 2D polarizability $\alpha_{2D}$ captures the macroscopic dielectric screening behavior of 2D materials [32]. Due to this weaker screening, 2D semiconductors and insulators have unusually large quasiparticle band gaps, exciton binding energies and band gap renormalizations compared with their bulk counterparts.

### III. GW self-energy and effective mass model: *h*-BN and MoS$_2$

As we can see from Eq. (1), to determine the quasiparticle self-energy of the doped system, the primary goal is to find the change in the dielectric screening, given by the dielectric function $\delta\epsilon_{GG'}^{-1}(\boldsymbol{q},\omega)$ of a 2D crystal. To illustrate this process in detail, we use *p*-doped *h*-BN as an example. *h*-BN is a wide-gap 2D insulator which has been commonly used as substrate and encapsulation for other 2D materials in Van der Waals heterostructures [33]. Its valence band maximum (VBM) is at the K point and conduction band minimum (CBM) at Γ point.

For a doped system, the change to the dielectric screening is concentrated on the head part of the dielectric function with small ***q*** and low frequency ω and requires a smaller number of bands to converge [23]. For this purpose, within the first-principles approach, the static dielectric function $\epsilon_{00}^{-1}(\boldsymbol{q},\omega=0)$ of the doped system is calculated on an $120 \times 120 \times 1$ k-point grid, as shown by the dots in Fig. 2. For the frequency-dependent part, a simple plasmon-pole model $\delta\epsilon_{00}^{-1}(\boldsymbol{q},\omega) = \frac{\delta\epsilon_{00}^{-1}(\boldsymbol{q},0)\omega_d^2(\boldsymbol{q})}{\omega^2-\omega_d^2(\boldsymbol{q})}$, where $\delta\epsilon_{00}^{-1}(\boldsymbol{q},0) = \epsilon_{00}^{-1}(\boldsymbol{q},0) - \epsilon_{int,00}^{-1}(\boldsymbol{q},0)$, well describes the difference between the intrinsic and doped dielectric function, and the plasmon frequency $\omega_d(\boldsymbol{q})$ is extracted from the *ab initio* calculation and shown in the inset of Fig. 2.

Following Ref. [23], the GW self-energy of the doped system can be calculated according to Eq. (1) term by term. The first correction term $\Sigma_1$ is given by

$$\Sigma_1^{nk}(E) = -\sum_{G,G'} \int \frac{d^2q}{(2\pi)^2} f_{n,k-q} M_{vn}^*(k,-q,-G) M_{vn}(k,-q,-G')$$
$$\times \epsilon_{int,GG'}^{-1}(q, E - \varepsilon_{n,k-q}) v_{2D}(q+G') \quad (2)$$

where $v$ is the doped band index, $f_{nk}$ is the electron occupation, $\varepsilon_{nk}$ is the mean-field (DFT) energy and $M_{nn'}(k,q,G)$ is the plane-wave matrix element. This self-energy is calculated from first-principle by taking the difference of the total self-energy of the intrinsic system from that of a doped one, both of which are evaluated with the dielectric function of the intrinsic system. To capture the change in occupation, the intrinsic dielectric function is calculated on a relatively dense k-point grid of $36 \times 36 \times 1$.

The other two terms $\Sigma_2$ and $\Sigma_3$ are expressed in summations that only involve intra-band transitions with small momentum as follow:

$$\Sigma_2^{nk}(E) = \pm \int \frac{d^2q}{(2\pi)^2} |M_{nn}(k,-q,0)|^2 \frac{\delta\epsilon_{00}^{-1}(q,0)}{2\left[1 \pm \frac{\varepsilon_{n,k-q}-E}{\omega_d(q)}\right]} v_{2D}(q) \quad (3)$$

$$\Sigma_3^{nk}(E) = -\int \frac{d^2q}{(2\pi)^2} \delta f_{n,k-q} |M_{nn}(k,-q,0)|^2 \frac{\delta\epsilon_{00}^{-1}(q,0)}{1-\left[\frac{\varepsilon_{n,k-q}-E}{\omega_d(q)}\right]^2} v_{2D}(q) \quad (4)$$

The $\pm$ in Eq. (3) is for conduction and valence states, respectively. Due to the interaction of the quasiparticle with the low-energy acoustic plasmon, $\Sigma_2$ and $\Sigma_3$ contains a resonance profile near the mean-field energy $\varepsilon_{nk}$. To this end, we employ the "on-shell" approximation to $\Sigma_2$ and $\Sigma_3$ by rigidly shifting the whole resonance profile along the energy axis such that the on-shell energy coincides with the QP solution [23]. The on-shell self-energy $\Sigma_1$, $\Sigma_2$ and $\Sigma_3$ of the VBM and CBM at K for p-doped h-BN calculated from first-principles are shown by the dots on Fig. 3.

However, this first-principles approach suffers a drawback as the dense k-point sampling required to accurately capture the electron occupation and dielectric screening limits its resolution at smaller doping density ($\sim 10^{12}/cm^{-2}$), which is, unfortunately, the most useful range for device applications. Therefore, we propose a first-principle-based effective mass model to solve this problem and gain insight for the band gap renormalization behavior at low doping density.

To construct the effective mass approximation for the dielectric function, we decompose the static polarizability function $\chi$ of the doped system as a sum of interband transitions and intraband transitions within the doped band. We assume the interband part remains the same as the polarizability of the intrinsic system, neglecting the small contributions from the virtual interband transitions near the VBM. The intraband part, within the effective mass approximation, is approximated by the polarizability of the two-dimensional electron gas (2DEG), given by the Lindhard function [34]:

$$\chi^{2DEG}(q, \omega = 0) = -\frac{N_s N_v m^*}{2\pi}[1 - \Theta(q - 2k_F)\sqrt{1 - \frac{4k_F^2}{q^2}}], \quad (5)$$

where $N_s = N_v = 2$ is the spin and valley degeneracy, $m^*$ is the effective mass of the 2DEG ($m^* = 0.78$ for $p$-doped $h$-BN), $k_F$ is the fermi wave vector and $\Theta$ is the step function. The singularity of $\chi^{2DEG}$ at $q = 2k_F$ manifests itself as a kink in the dielectric function, as indicated by the arrow in Fig. 2.

Given the assumptions above, the static polarizability within the effective mass model is $\chi_{GG'}(q, 0) = \chi_{GG'}^{int}(q, 0) + \frac{1}{L}\chi^{2DEG}(q, 0)$ for all G-vectors with $G_x=G_y=0$, where $L$ is the cell periodicity in the z-direction. The RPA dielectric function is then determined by $\epsilon_{GG'}(q, 0) = \delta_{GG'} - \chi_{GG'}(q, 0)v_{2D}(q + G')$, where $v_{2D}(q) = \frac{4\pi}{q^2}[1 - e^{-q_{xy}L/2}\cos(\frac{q_zL}{2})]$ is the 2D truncated Coulomb interaction [28]. The input from *ab initio* calculations can be further reduced by observing that the behavior of the intrinsic polarizability $\chi_{GG'}^{int}(q, 0)$ as $q \to 0$ is determined by the 2D polarizability: $\chi_{GG'}^{int}(q, 0) = \chi_{GG'}^{int}(0,0) - \frac{\alpha_{2D}}{L}q^2$. In practice, we find that only including the $G_z=0, \pm1$ elements of $\chi_{GG'}^{int}(0,0)$ is sufficient to construct an effective mass model for $\epsilon_{00}^{-1}(q, 0)$ that accurately reproduces the *ab initio* one, as shown by the lines in Fig. 2. Meanwhile, within the effective mass approximation, the plasmon-pole frequency follows the 2DEG dispersion relation $\omega_d^{2DEG}(q) = \sqrt{\frac{2\pi n q}{m}\left(1 + \frac{q}{2}\right)^2\left(1 + \frac{q^3}{8\pi n} + \frac{q^4}{32\pi n}\right)/(1 + \frac{q}{4})}$ [35], which also fits the *ab initio* values well, as shown by the inset of Fig. 2.

With the effective mass model, we calculate asymptotic behavior of self-energy terms Eq. (2)-(4) in the low density limit. At low doping density, keeping only the leading contribution, $\Sigma_1$ at the VBM reduces to

$$\Sigma_1^{VBM} \sim \int_{q<k_F} \frac{d^2q}{(2\pi)^2}\epsilon_{int,00}^{-1}(q, 0)v_{2D}(q). \quad (6)$$

Meanwhile, the on-shell self-energy $\Sigma_2$ and $\Sigma_3$ are reduced to the following as $q \to 0$:

$$\Sigma_2^{VBM} \sim -\frac{1}{2}\int \frac{d^2q}{(2\pi)^2} \frac{\delta\epsilon_{00}^{-1}(q,0)}{1-\varepsilon_q/\omega_d(q)} v_{2D}(q), \tag{7}$$

$$\Sigma_3^{VBM} \sim \int_{q<k_F} \frac{d^2q}{(2\pi)^2} \delta\epsilon_{00}^{-1}(q,0) v_{2D}(q), \tag{8}$$

where the term $\varepsilon_q/\omega_d(q)$ is dropped from Eq. (4) because as $q \to 0$, $\varepsilon_q \propto q^2$ while $\omega_d(q) \propto \sqrt{q}$ so $\varepsilon_q/\omega_d(q) \to 0$. In the leading order, both $\Sigma_1$ and $\Sigma_3$ affect only the band which has been doped (and does not affect the self-energy at the CBM), while $\Sigma_2$ affects all states at the same time.

Equations (6) and (8) share a similar form of an integral over the doped region. Equation (6) shows that $\Sigma_1$ correspond to "bare" exchange energy of a 2DEG, where the bare interaction refers to the screened interaction of the intrinsic system without the additional screening from the 2DEG. Meanwhile, Eq. (8) suggests that $\Sigma_3$ correspond the difference between the "bare" exchange and the screened exchange energy of 2DEG. In fact, $\Sigma_3$ cancels most part of $\Sigma_1$, because $\epsilon_{int,00}^{-1}(q,0) \gg \epsilon_{00}^{-1}(q,0)$ for $q < k_F$ and thus $\Sigma_1 \gg \Sigma_1 + \Sigma_3$. Their sum

$$\Sigma_1^{VBM} + \Sigma_3^{VBM} \sim \int_{q<k_F} \frac{d^2q}{(2\pi)^2} \epsilon_{00}^{-1}(q,0) v_{2D}(q), \tag{9}$$

is the actual screened exchange contribution to the self-energy. It grows linearly with the doping density because $\epsilon_{00}^{-1}(q,0)$ is linear in q as $q \to 0$. Due to the 2DEG polarizability from Eq. (5), it is also proportional to inverse of the density-of-state effective mass $1/N_s N_v m^*$. The linear behavior from this asymptotic analysis, as shown by the red line from Fig. 3, accurately describes the *ab initio* results, even for the points with relatively high doping density.

On the other hand, $\Sigma_2$, which corresponds to the Coulomb-hole part of the self-energy [36], has a very different asymptotic behavior at low doping density. The integral in Eq. (7) goes over the whole BZ. As the integrant, the change in dielectric function $\delta\epsilon_{00}^{-1}(q,0)$, given by the difference between the curves in Fig. 2, is rapidly increasing at low doping density but saturates at high doping density. This causes the term $\Sigma_2$ to dominate the low-density part of the band gap renormalization, and saturate at high density. The self-energy calculated from Eq. (7) is shown by the black and blue curves in Fig. 3 and they are also in good agreement with the *ab initio* results. To sum up, *it is shown that the band gap renormalization is dominated by the nonlinear Coulomb-hole term ($\Sigma_2$) in the low doping density region, while the linearly increasing screened exchange term ($\Sigma_1 + \Sigma_3$) takes over in the high doping density region as the Coulomb-hole term saturates.*

Finally, we show the quasiparticle band gap renormalization of *p*-doped *h*-BN in Fig. 4. Based on our calculated DFT and GW band structure shown in Fig. 4 (a), intrinsic *h*-BN has an indirect band gap of 6.4 eV with VBM at the K point and CBM at Γ point of the Brillouin zone. The direct band gap at K is 7.3 eV. Fig. 4 (b) shows the renormalization of the direct band gap at K. With hole doping, the band gap drops rapidly by about 1 eV with doping density around $10^{12}$-$10^{13}$cm$^{-2}$. With further increase in doping density, the band gap renormalization saturates to a slower rate. The renormalizations of the VBM and CBM quasiparticle energy are shown in the inset of Fig. 4 (b). They are nearly symmetric because the dominating Coulomb-hole self-energy term given by Eq. (7), which is not sensitive to which band is occupied by doped carriers, makes almost equal but opposite contribution to valence and conduction band. The slight difference is from the fact that the screened exchange term affects the doped band, causing the VBM energy to have a larger shift than the CBM at large doping density.

In Fig. 5, we show similar results for the *n*-doped MoS$_2$. Despite having a much smaller intrinsic band gap around 2.7 eV (without considering the spin-orbit coupling), MoS$_2$ shares similar honeycomb lattice structure and isotropic effective mass with *h*-BN. Therefore, MoS$_2$ shows a similar band gap renormalization behavior. A moderate doping density (around $10^{13}$ cm$^{-2}$) can induce a band gap reduction of 400 meV. The solid line is from our effective mass model. It perfectly captures the low-density results while slightly overestimates the reduction for high doping densities. This is not surprising because our effective mass model does not include the band structure effects and the off-diagonal elements of the dielectric function, which would gradually gain importance at higher doping density.

## IV.     Band gap renormalization of monolayer BP

BP is a layered semiconductor that has attracted great interest recently [20, 37, 38]. It has a direct band gap that is tunable with the number of layers, ranging from 0.3 eV in bulk to 2.0 eV in a monolayer [39]. Adatoms and doping have been found to strongly affect the band gap of thin-film BP [40]. It also shows strong in-plane anisotropy, which results in unusual behaviors of anisotropic exciton and thermal and electrical transport [41, 42]. The band structure of monolayer BP is shown in Fig. 7 (a). The most special character is that, near the band edge at Γ point, BP has a parabolic band dispersion with large effective mass in the x (zigzag)-direction and an almost linear band dispersion with very small effective mass in the y (armchair)-direction. Consequently, the screening in intrinsic and doped BP are also anisotropic. Therefore, we must modify the above isotropic effective mass model to calculate the band gap renormalization in doped monolayer BP.

The static dielectric function $\epsilon_{00}^{-1}(\boldsymbol{q}, \omega = 0)$ of intrinsic and doped BP is calculated on a $112 \times 80 \times 1$ k-point grid and their values along the x- and y-directions are shown in Fig. 6 (a), respectively. It is clear that the dielectric screening of both the intrinsic and doped system are anisotropic. Notably the kink at $q = 2k_F$ due to the singularity in the 2DEG polarizability is still present in the dielectric function of doped BP, although $k_F$ takes different values in x and y directions. Before the kink $\epsilon^{-1}$ is isotropic and corresponds to a constant polarizability of the 2DEG despite its anisotropic effective mass, while after the kink $\epsilon^{-1}$ turns up and merges into the intrinsic dielectric function. It should be noted that although the effective mass along x- and y-directions differ by about a factor of 7, the difference of the intrinsic and doped dielectric function is only weakly dependent on the direction of $\boldsymbol{q}$.

Contrast to the static case, the band anisotropy has a much greater impact on the frequency-dependent part of the dielectric function. The polar plot in Fig. 6 (b) shows the loss function $Im[\epsilon_{00}^{-1}(\boldsymbol{q},\omega)]$ as a function of $\omega$ and the direction of $\boldsymbol{q}$. The darker region in the plot corresponds to a peak in the loss function corresponding to the plasmon excitation, showing that the plasmon is highly anisotropic in BP. We find that the angular-dependent plasmon frequency can be well fitted by the relation $\omega_d(\boldsymbol{q}) \propto \sqrt{\frac{\cos^2\theta}{m_x} + \frac{\sin^2\theta}{m_y}}$, where $m_x = 1.22 m_0$ and $m_y = 0.16 m_0$ are the electron effective masses in the two directions and $\theta$ is the direction of $\boldsymbol{q}$. Apart from this anisotropy, the plasmon frequency is follows the characteristic of 2DEG and is proportional to $\sqrt{q}$ and $\sqrt{n}$ for small $q$ and low doping density $n$. The screening properties of BP obtained with our *ab initio* calculation agree well with a previous study using the effective Hamiltonian approach [43].

The quasiparticle self-energy of the doped BP is expanded similarly into $\Sigma_1$, $\Sigma_2$ and $\Sigma_3$ following Eq. (1). Each term is calculated according to Eq. (2)-(4) with the difference that the integral over **q** now needs to be done in 2D instead of 1D. The *ab initio* static dielectric function $\epsilon_{00}^{-1}(\boldsymbol{q},\omega=0)$ and the plasmon frequency $\omega_d(\boldsymbol{q})$ on the 2D grid is used as input for the integrals. We find these two-dimensional integrals can be further simplified by modelling the angular dependence of $\epsilon_{00}^{-1}(\boldsymbol{q},\omega=0)$ and $\omega_d(\boldsymbol{q})$. By assuming $\delta\epsilon_{00}^{-1}(\boldsymbol{q},\omega=0)$ to be the average of x- and y-direction and isotropic, as well as using the angular dependence of $\omega_d(\boldsymbol{q})$ shown above, we can further reduce the q-points needed to for the *ab initio* calculation to only along the line Γ-X and Γ-Y. This yields similar result to the full 2D integration with a difference in the on-shell self-energy at VBM and CBM of less than 10meV.

The resulting quasiparticle band gap renormalization of *n*-doped BP is shown in Fig. 7 (b). The quasiparticle band gap drops rapidly from 1.95eV to around 1.58eV with light doping up to density n = 2×10$^{12}$cm$^{-2}$. However, there is a notable difference from *h*-BN and MoS$_2$ that the band gap renormalization of BP is less saturated at high doping density. As the inset in Fig. 7 (b) shows, this is due to a continued decrease of the CBM quasiparticle energy at large doping density, while the VBM quasiparticle energy has already saturated to nearly constant. The on-shell self-energy values at VBM and CBM, shown in Figs. 7 (c) and (d), reveal the reason behind this unusual behavior. Same as *h*-BN and MoS$_2$, the Coulomb-hole term $\Sigma_2$, as shown by the red curves, is dominant at low doping density but saturates at higher doping density. However, the screened-exchange term $\Sigma_1+\Sigma_3$ as shown by the magenta curve in Fig. 7 (d), which controls the CBM self-energy renormalization at high density, is notably larger than that in *h*-BN and MoS$_2$. As we have discussed in the asymptotic analysis, the screened-exchange self-energy is inversely proportional to the density-of-state effective mass. Due to the lack of valley degeneracy and highly anisotropic, quasi-1D band dispersion, electrons in BP has a small DOS effective mass $\sqrt{m_x m_y} \approx 0.44$, which is about 2 times smaller than MoS$_2$ and 3 times smaller than *h*-BN. The calculated slope of the screened-exchange self-energy versus doping density is indeed 3 times larger for BP than *h*-BN, which confirms that the smaller DOS of BP is the root cause of its large, unsaturated band gap renormalization.

## V.     Summary

In summary, we have discussed the band gap renormalization in doped 2D materials within the GW approximation for three prototypical materials: *h*-BN and MoS$_2$ and black phosphorus. We have combined *ab initio* results and effective mass model to determine the dielectric screening, quasiparticle self-energy and band gap renormalization at arbitrary doping density. With asymptotic analysis, we have shown that the main contribution to the band gap renormalization can be separated into two terms. One is the Coulomb-hole term coming from the change of the dielectric screening, which is highly nonlinear and dominant at low doping density. The other is the screened-exchange term coming from the change in electron occupation, which is linear and more important at high doping density. We have also studied the anisotropic dielectric screening of BP. We find BP has a larger band gap renormalization at high doping density, which we attribute to the smaller density-of-state of BP near the band edge.

**ACKNOWLEDGMENT**

Authors are supported by the National Science Foundation (NSF) CAREER Grant No. DMR-1455346 and NSF EFRI-2DARE-1542815. The computational resources have been provided by the Stampede of Teragrid at the Texas Advanced Computing Center (TACC) through XSEDE.

**Figures:**

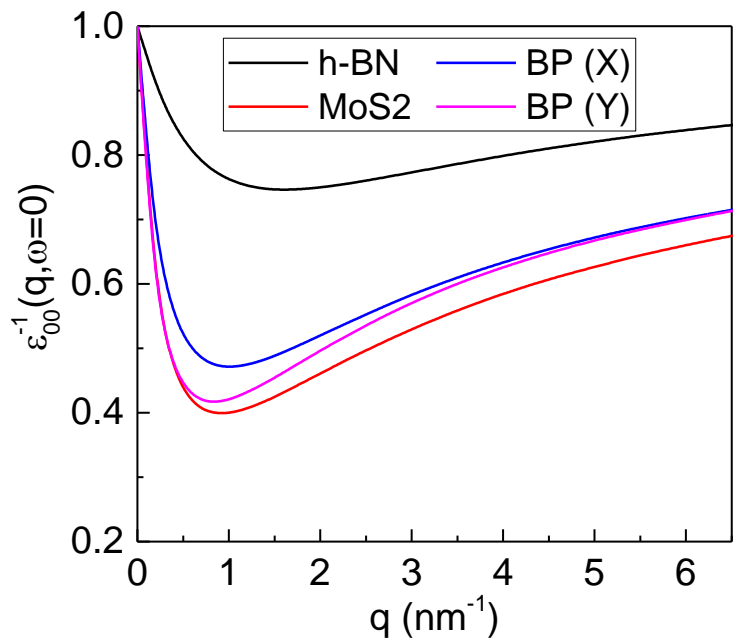

Fig. 1 Static dielectric function $\epsilon_{00}^{-1}(\boldsymbol{q}, \omega = 0)$ of intrinsic *h*-BN, MoS$_2$ and BP with the same size of vacuum (20Å).

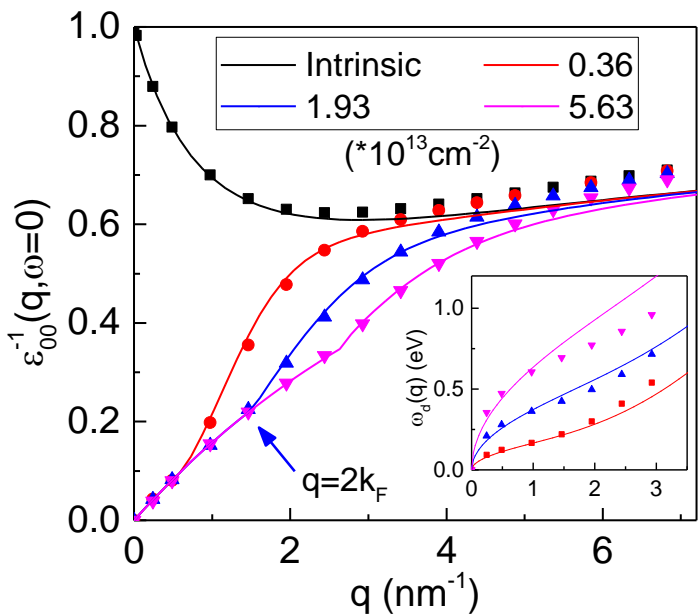

Fig. 2 Static dielectric function of *p*-doped *h*-BN. Dots are from the *ab initio* calculation and the solid

lines come from the effective mass model. The inset shows the plasmon-pole frequency.

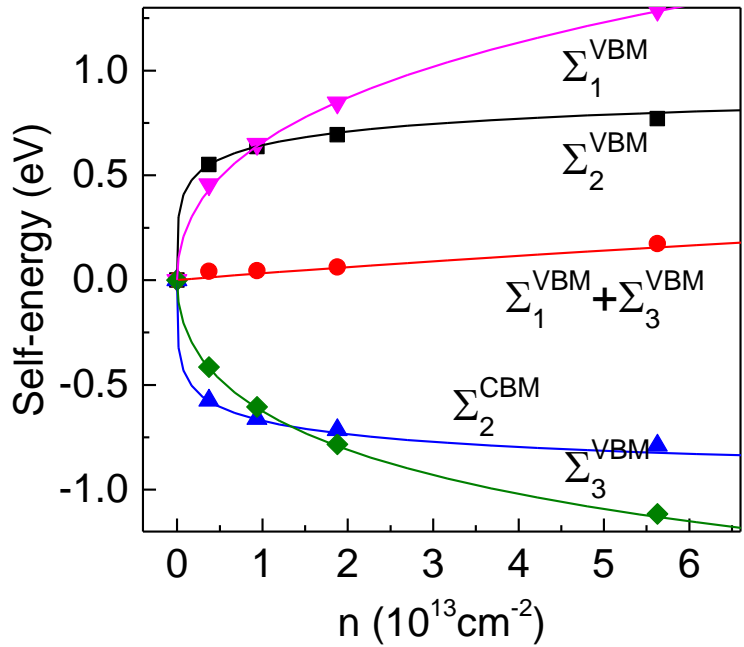

Fig. 3 On-shell self-energy of *p*-doped *h*-BN at VBM and CBM. Dots represent the *ab initio* result and the solid line is from the effective mass model.

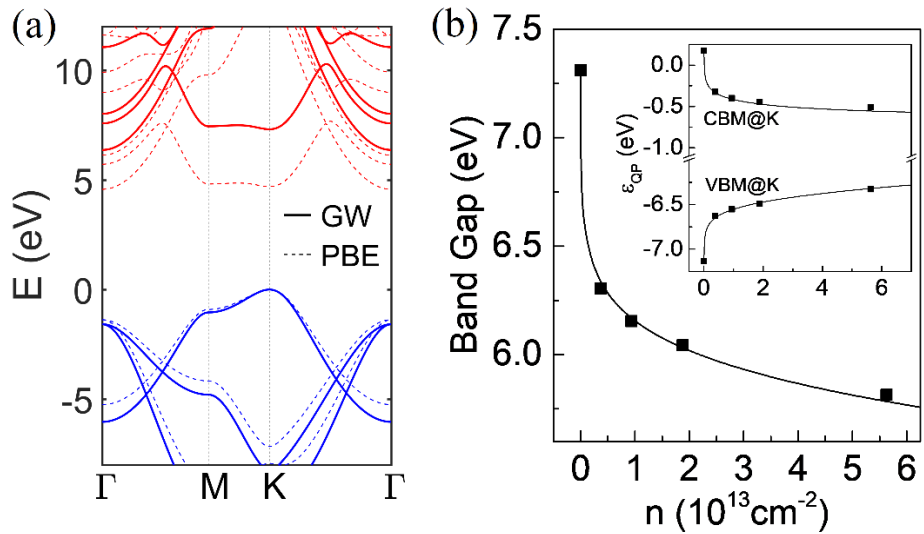

Fig. 4 (a) DFT and GW band structure of intrinsic *h*-BN. (b) Renormalization of the direct band gap at K for *p*-doped *h*-BN. Inset shows the quasiparticle energy. Dots represent the *ab initio* result and the solid line is from the effective mass model.

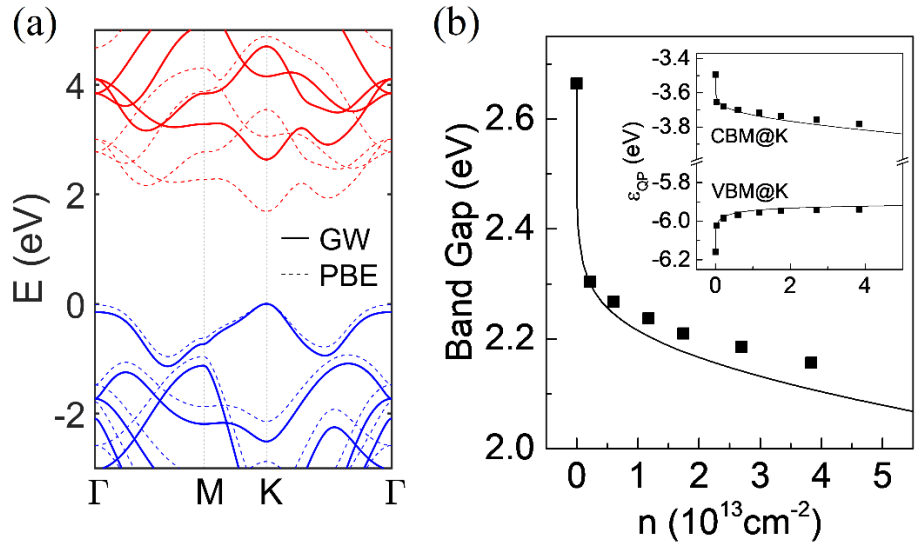

Fig. 5 (a) DFT and GW band structure of intrinsic MoS$_2$. (b) Quasiparticle band gap renormalization of *n*-doped MoS$_2$. Dots represent the *ab initio* result and the solid line is from the effective mass model.

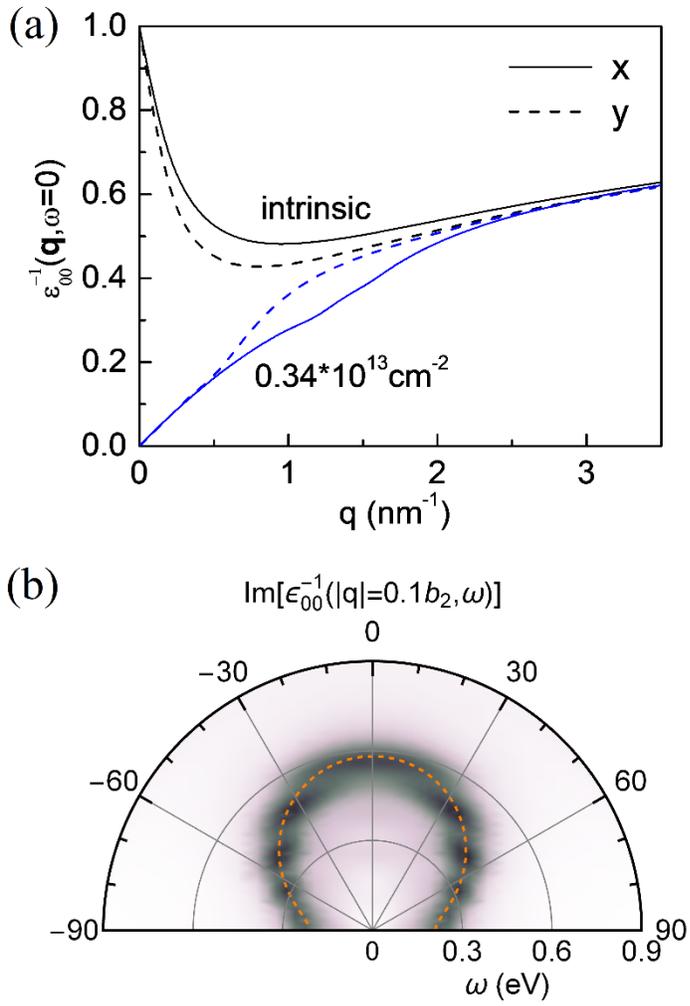

Fig. 6 (a) First-principles static dielectric function of *n*-doped BP in different direction. (b) Polar plot of the loss function in *n*-doped BP. The dashed line is a fit to the plasmon frequency with the anisotropic effective mass.

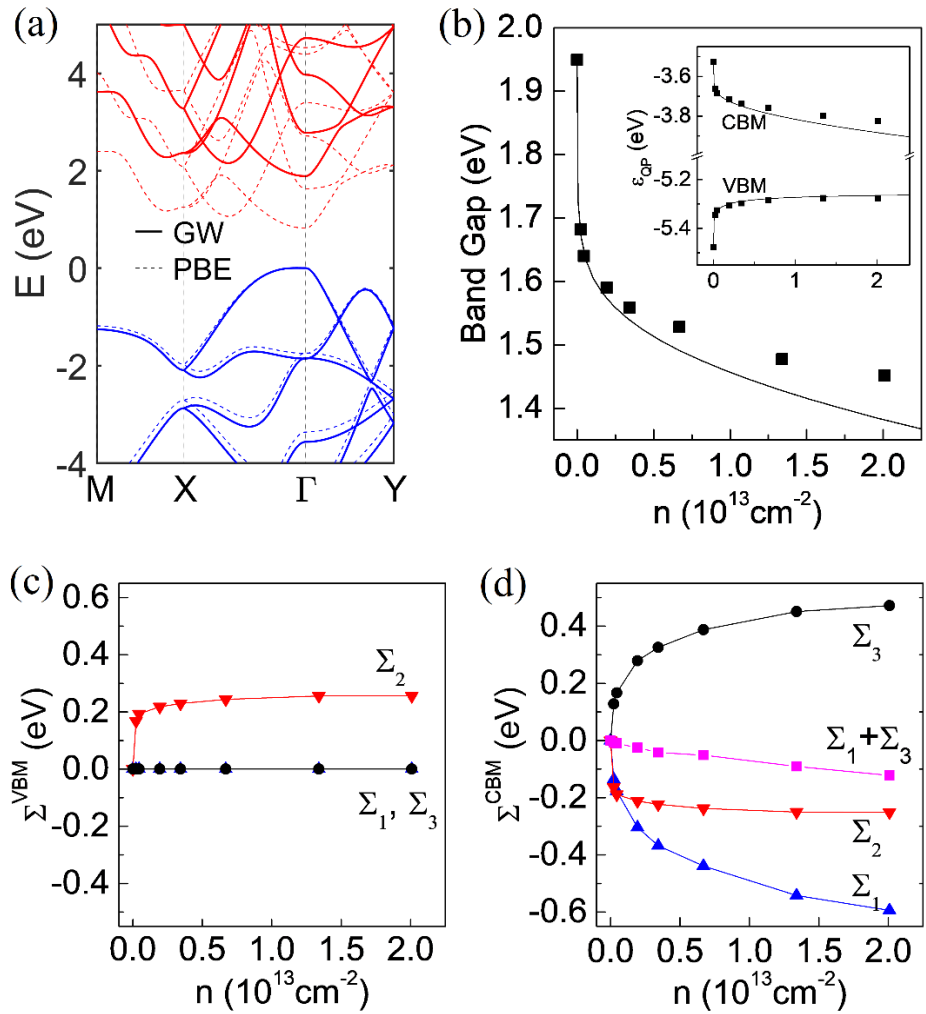

Fig. 7 (a) DFT and GW band structure of intrinsic BP. (b) Quasiparticle band gap renormalization of *n*-doped BP. Dots represent the *ab initio* result and the solid lines are from the effective mass model. (c) (d) The on-shell self-energy $\Sigma_1$, $\Sigma_2$ and $\Sigma_3$ at the VBM and CBM as a function of doping density.

**REFERENCES**

[1] Q. H. Wang, Q. Hua, K. Kalantar-Zadeh, A. Kis, J. N. Coleman, and M. S. Strano, Electronics and optoelectronics of two-dimensional transition metal dichalcogenides, Nature Nanotechnology **7**, 699-712 (2012).

[2] M. Bernardi, M. Palummo, and J. C. Grossman, Extraordinary sunlight absorption and one nanometer thick photovoltaics using two-dimensional monolayer materials, Nano Letters **13**, 3664-3670 (2013).

[3] L. Britnell, R. M. Ribeiro, A. Eckmann, R. Jalil, B. D. Belle, A. Mishchenko, Y.-J. Kim, R. V. Gorbachev, T. Georgiou, S. V. Morozov, A. N. Grigorenko, A. K. Geim, C. Casiraghi, A. H. Castro Neto, K. S. Novoselov, Strong light-matter interactions in heterostructures of atomically thin films, Science **340**, 1311-1314 (2013).

[4] H. Tian, J. Tice, R. Fei, V. Tran, X. Yan, L. Yang, and H. Wang, Low-symmetry two-dimensional materials for electronic and photonic applications, Nano Today **11**, 763-777 (2016).

[5] R. Fei, W. Kang, and L. Yang, Ferroelectricity and Phase Transitions in Monolayer Group-IV Monochalcogenides, Phys. Rev. Lett. **117**, 097601 (2016).

[6] Cheng Gong, Lin Li, Zhenglu Li, Huiwen Ji, Alex Stern, Yang Xia, Ting Cao, Wei Bao, Chenzhe Wang, Yuan Wang, Z. Q. Qiu, R. J. Cava, Steven G. Louie, Jing Xia, and Xiang Zhang, Discovery of intrinsic ferromagnetism in two-dimensional van der Waals crystals, Nature **546**, 265-269 (2017).

[7] Bevin Huang, Genevieve Clark, Efrén Navarro-Moratalla, Dahlia R. Klein, Ran Cheng, Kyle L. Seyler, Ding Zhong, Emma Schmidgall, Michael A. McGuire, David H. Cobden, Wang Yao, Di Xiao, Pablo Jarillo-Herrero, and Xiaodong Xu, Layer-dependent ferromagnetism in a van der Waals crystal down to the monolayer limit, Nature **546**, 270-273 (2017).

[8] F. Wu, F. Xue, and A. H. MacDonald, Theory of two-dimensional spatially indirect equilibrium exciton condensates, Phys. Rev. B **92**, 165121 (2015).

[9] G. Onida, L. Reining, and A. Rubio, Electronic excitations: density-functional versus many-body Green's-function approaches, Rev. Mod. Phys. **74**, 601 (2002).

[10]     M. S. Hybertsen and S. G. Louie, Electron correlation in semiconductors and insulators: Band gaps and quasiparticle energies, Phys. Rev. B **34**, 5390 (1986).

[11]     Jack Deslippe, Georgy Samsonidze, David A Strubbe, Manish Jain, Marvin L Cohen, and Steven G Louie, BerkeleyGW: A massively parallel computer package for the calculation of the quasiparticle and optical properties of materials and nanostructures, Comp. Phys. Comm. **183**, 1269-1289 (2012).

[12]     C. D. Spataru, S. Ismail-Beigi, L. X. Benedict, and S. G. Louie, Excitonic effects and optical spectra of single-walled carbon nanotubes, Phys. Rev. Lett. **92**, 077402 (2004).

[13]     F. Hüser, T. Olsen, and K. S. Thygesen, How dielectric screening in two-dimensional crystals affects the convergence of excited-state calculations: Monolayer $MoS_2$, Phys. Rev. B **88**, 245309 (2013).

[14]     F. A. Rasmussen, P. S. Schmidt, K. T. Winther, and K. S. Thygesen, Efficient many-body calculations for two-dimensional materials using exact limits for the screened potential: Band gaps of $MoS_2$, *h*-BN, and phosphorene. Phys. Rev. B **94**, 155406 (2016).

[15]     D. Y. Qiu, H. Felipe, and S. G. Louie, Screening and many-body effects in two-dimensional crystals: Monolayer $MoS_2$, Phys. Rev. B **93**, 235435 (2016).

[16]     H. Felipe, D. Y. Qiu, and S. G. Louie, Nonuniform sampling schemes of the Brillouin zone for many-electron perturbation-theory calculations in reduced dimensionality, Phys. Rev. B **95**, 035109 (2017).

[17]     K. F. Mak, K. He, C. Lee, G. H. Lee, J. Hone, T. F. Heinz, and J. Shan, Tightly bound trions in monolayer $MoS_2$, Nature Materials **12**, 207-211 (2013).

[18]     S. Mouri, Y. Miyauchi, and K. Matsuda, Tunable photoluminescence of monolayer $MoS_2$ via chemical doping, Nano Letters **13**, 5944-5948 (2013).

[19]     Yi Zhang, Tay-Rong Chang, Bo Zhou, Yong-Tao Cui, Hao Yan, Zhongkai Liu, Felix Schmitt, James Lee, Rob Moore, Yulin Chen, Hsin Lin, Horng-Tay Jeng, Sung-Kwan Mo, Zahid Hussain, Arun Bansil, and Zhi-Xun Shen, Direct observation of the transition from indirect to direct bandgap in atomically thin epitaxial $MoSe_2$, Nature Nanotechnology **9**, 111-115 (2014).

[20] L. Li, Y. Yu, G. J. Ye, Q. Ge, X. Ou, H. Wu, D. Feng, X. H. Chen, and Y. Zhang, Black phosphorus field-effect transistors, Nature Nanotechnology **9**, 372-377 (2014).

[21] C. D. Spataru, and F. Léonard, Tunable band gaps and excitons in doped semiconducting carbon nanotubes made possible by acoustic plasmons, Phys. Rev. Lett. **104**, 177402 (2010).

[22] C. D. Spataru, and F. Léonard, Quasiparticle and exciton renormalization effects in electrostatically doped semiconducting carbon nanotubes, Chemical Physics **413**, 81-88 (2013).

[23] Y. Liang, and L. Yang, Carrier plasmon induced nonlinear band gap renormalization in two-dimensional semiconductors, Phys. Rev. Lett. **114**, 063001 (2015).

[24] S. Larentis, J. R. Tolsma, B. Fallahazad, D. C. Dillen, K. Kim, A. H. MacDonald, and E. Tutuc, Band offset and negative compressibility in graphene-$MoS_2$ heterostructures, Nano Letters **14**, 2039-2045 (2014).

[25] J. M. Riley, W. Meevasana, L. Bawden, M. Asakawa, T. Takayama, T. Eknapakul, T. K. Kim, M. Hoesch, S.-K. Mo, H. Takagi, T. Sasagawa, M. S. Bahramy, and P. D. C. King, Negative electronic compressibility and tunable spin splitting in $WSe_2$, Nature Nanotechnology **10**, 1043-1047 (2015).

[26] A. Chernikov, C. Ruppert, H. M. Hill, A. F. Rigosi, and T. F. Heinz, Population inversion and giant bandgap renormalization in atomically thin $WS_2$ layers, Nature Photonics **9**, 466-470 (2015).

[27] Eva A. A. Pogna, Margherita Marsili, Domenico De Fazio, Stefano Dal Conte†, Cristian Manzoni, Davide Sangalli, Duhee Yoon, Antonio Lombardo, Andrea C. Ferrari, Andrea Marini, Giulio Cerullo, and Deborah Prezzi, Photo-induced bandgap renormalization governs the ultrafast response of single-layer $MoS_2$, ACS Nano **10**, 1182-1188 (2016).

[28] S. Gao, Y. Liang, C. D. Spataru, and L. Yang, Dynamical Excitonic Effects in Doped Two-Dimensional Semiconductors, Nano Letters **16**, 5568-5573 (2016).

[29] P. Giannozzi, et al., QUANTUM ESPRESSO: a modular and open-source software project for quantum simulations of materials, Journal of physics: Condensed matter **21**, 395502 (2009).

[30] J. P. Perdew, K. Burke, and M. Ernzerhof, Generalized gradient approximation made simple, Phys. Rev. Lett. **77**, 3865 (1996).

[31] S. Ismail-Beigi, Truncation of periodic image interactions for confined systems, Phys. Rev. B **73**, 233103 (2006).

[32] P. Cudazzo, I. V. Tokatly, and A. Rubio, Dielectric screening in two-dimensional insulators: Implications for excitonic and impurity states in graphene, Phys. Rev. B **84,** 085406 (2011).

[33] A. K. Geim, and I. V. Grigorieva, Van der Waals heterostructures, Nature **499**, 419-425 (2013).

[34] G. F. Giuliani, and G. Vignale, Quantum Theory of the Electron Liquid, Cambridge University Press (2005).

[35] A. Czachor, A. Holas, S. R. Sharma, and K. S. Singwi, Dynamical correlations in a two-dimensional electron gas: First-order perturbation theory, Phys. Rev. B **25**, 2144 (1982).

[36] L. Hedin, New method for calculating the one-particle Green's function with application to the electron-gas problem, Phys. Rev. **139**, A796 (1965).

[37] H. Liu, A. T. Neal, Z. Zhu, Z. Luo, X. Xu, D. Tománek, and D. Y. Peide, Phosphorene: an unexplored 2D semiconductor with a high hole mobility, ACS Nano **8**, 4033-4041 (2014).

[38] F. Xia, H. Wang, and Y. Jia, Rediscovering black phosphorus as an anisotropic layered material for optoelectronics and electronics, Nature Communications **5**, 4458 (2014).

[39] V. Tran, R. Soklaski, Y. Liang, and L. Yang, Layer-controlled band gap and anisotropic excitons in few-layer black phosphorus, Phys. Rev. B **89**, 235319 (2014).

[40] Jimin Kim, Seung Su Baik, Sae Hee Ryu, Yeongsup Sohn, Soohyung Park, Byeong-Gyu Park, Jonathan Denlinger, Yeonjin Yi, Hyoung Joon Choi, and Keun Su Kim, Observation of tunable band gap and anisotropic Dirac semimetal state in black phosphorus, Science **349**, 723-726 (2015).

[41] R. Fei, and L. Yang, Strain-engineering the anisotropic electrical conductance of few-layer black phosphorus, Nano Letters **14**, 2884-2889 (2014).

[42] R. Fei, A. Faghaninia, R. Soklaski, J. Yan, C. Lo, and L. Yang, Enhanced thermoelectric efficiency via orthogonal electrical and thermal conductances in phosphorene, Nano Letters **14**, 6393-6399 (2014).

[43]    T. Low, R. Roldán, H. Wang, F. Xia, P. Avouris, L. M. Moreno, and F. Guinea, Plasmons and screening in monolayer and multilayer black phosphorus, Phys. Rev. Lett. **113**, 106802 (2014).